\documentclass[12pt]{iopart}
\usepackage{graphicx}
\usepackage{epstopdf}
\begin{document}
\title{Remarks on the interquark potential in the presence of a minimal length}
\author{Patricio Gaete\dag\footnote{e-mail address: patricio.gaete@usm.cl}}
\address{\dag\ Departamento de F\'{\i}sica and Centro Cient\'{i}fico-Tecnol\'ogico de Valpara\'{i}so, Universidad T\'ecnica Federico Santa Mar\'{\i}a, Valpara\'{\i}so, Chile}

\begin{abstract}
We calculate the lowest-order corrections to the static potential for both Yang-Mills theory and gluodynamics in curved space-time, in the presence of a quantum gravity induced minimal length. Our analysis is carried out within stationary perturbation theory. As a consequence, the potential energy is the sum of a Yukawa-like potential and a linear potential for gluodynamics in curved space-time, leading to the confinement of static charges. Interestingly enough, we obtain that the coefficient of the linear term ("string tension") is ultraviolet finite. We should highlight the role played by the new quantum of length in our analysis.
\end{abstract}
\submitted
\maketitle

\section{Introduction}

The study of extensions of the Standard Model (SM), such as Lorentz invariance violation and fundamental length, have motivated much research effort  in the past years \cite{AmelinoCamelia:2002wr,Jacobson:2002hd,Konopka:2002tt,Hossenfelder:2006cw,Nicolini:2008aj}. Clearly, such studies are supported because the SM does not include a quantum theory of gravitation. In fact, the necessity of a new scenario has been suggested to overcome difficulties theoretical in the quantum gravity research. In this respect we recall that string theories \cite{Witten} provide a consistent framework to unify all fundamental interactions. 
Let us also mention here that during the last years the focus of quantum gravity has been towards effective models, which have helped to gain insights on a new and unconventional physics. Actually, these models incorporate one of the most important and general features, that is, a \emph{minimal length scale} that acts as a regulator in the ultraviolet. Mention should be made, at this point, to different approaches which incorporate a minimal length \cite{Magueijo:2001cr,Hossenfelder:2003jz,Nicolini:2010bj,Sprenger:2012uc}. Among these researches, probably the most studied framework are quantum field theories allowing non-commuting position operators 
\cite{Witten:1985cc,Seiberg:1999vs,Douglas:2001ba,Szabo:2001kg,Gomis:2000sp,Bichl:2001nf}. 

Another interesting observation is that most of the results in the
non-commutative approach have been achieved using the so called
star-product (Moyal product). However, as it is known, there is another formulation of  non-commutative quantum field theory. This alternative formulation is also known as quantum field theory in the presence of a minimal length, which has been proposed in \cite{Smailagic:2003rp,Smailagic:2003yb,Smailagic:2004yy}. Basically, this approach is based on defining the fields as mean value
over coherent states of the noncommutative plane, such that a star
product needs not be introduced. Subsequently, it has been shown
that the coherent state approach can be summarized through the
introduction of a new multiplication rule which is known as Voros
star-product \cite{Galluccio:2008wk,Galluccio:2009ss},
\cite{Banerjee:2009xx,Gangopadhyay:2010zm,Basu:2011kh}. Anyway,
physics turns out be independent from the choice of the type of
product \cite{Hammou:2001cc}. An alternative view of these
modifications  is to consider them as a redefinition of the Fourier
transform of the fields. Accordingly, the theory is ultraviolet
finite and the cutoff is provided by the noncommutative parameter
$\theta$. Notice that the existence of a minimal length is determined by the noncommutative parameter $\theta$. In a general perspective, since one can incorporate a minimal length ($\sqrt{\theta}$) in spacetime by assuming nontrivial coordinate commutation relations, we then have introduced a noncommutative geometry. 

With these ideas in mind, in previous studies
\cite{Gaete:2011ka,Gaete:2012yu}, we have considered the effect of the spacetime noncommutativity on a physical observable. In fact, we have computed the static potential for axionic electrodynamics both in $(3+1)$ and  $(2+1)$ space-time dimensions, in the presence of a minimal length. The point we wish to emphasize, however, is that our analysis leads to a well-defined noncommutative interaction energy. 
Indeed, in both cases we have obtained a fully ultraviolet finite static potential. Seen from such a perspective, the purpose of the present work is to extend the Abelian calculations  to the non-Abelian case and find the corresponding static potential. Our calculations are done by using perturbation theory along the lines of Refs. \cite{Drell:1981gu,Gribov:1977ja,Bagan:2000nc}. One important advantage of this approach is that it provides a transparent description of the origin of asymptotic freedom in gauge theories. In short, we will calculate the lowest-order corrections to the static potential for both Yang-Mills theory and gluodynamics in curved space-time, in the presence of a minimal length. 

\section{Yang-Mills in the presence of a minimal length}  

As stated in the introduction, the main focus of this paper
is to reexamine the interaction energy for gluodynamics
in curved space-time in the presence of a minimal length. 
However, let us start our considerations by considering a 
noncommutative version of Yang-Mills theory to illustrate the
central features and will then be extended to noncommutative
gluodynamics. Specifically, in this work we will focus attention on the short distance perturbative interaction potential.

To this end, the initial point of our analysis is the following four-dimensional space-time Lagrangian:
\begin{eqnarray}
{\cal L} =  - \frac{1}{4}Tr\left( {F_{\mu \nu } F^{\mu \nu } } \right) 
=  - \frac{1}{4}F_{\mu \nu }^a F^{a\mu \nu },
\label{NCYM05}
\end{eqnarray}
where $ A_\mu  \left( x \right) = A_\mu ^a \left( x \right)T^a $ and
$ F_{\mu \nu }^a  = \partial _\mu  A_\nu
^a - \partial _\nu  A_\mu ^a  + gf^{abc} A_\mu ^b A_\nu ^c$, with
$f^{abc}$ the structure constants of the gauge group. 

As already mentioned, our analysis is based in perturbation theory along the lines of Refs. \cite{Drell:1981gu,Gribov:1977ja,Bagan:2000nc}. In other words, we will work out the vacuum expectation value of the energy operator $H$ $(\left\langle 0 \right|H\left| 0 \right\rangle)$ at lowest order in $g$. In such a case, it is profitable to use the Coulomb gauge because the fields have a simple physical meaning. Thus the Hamiltonian corresponding to (\ref{NCYM05})  reads
\begin{equation}
H = \frac{1}{2}\int {{d^3}x} \left\{ {{{\left( {{\bf E}_T^a} \right)}^2} +
{{\left( {{{\bf B}^a}} \right)}^2} - {\phi ^a}{\nabla ^2}{\phi ^a}} \right\}, 
\label{NCYM10}
\end{equation}
where the color-electric field ${{\bf E}^a }$ has been separated into transverse and longitudinal parts: ${{\bf E}^a} = {\bf E}_T^a - \nabla {\phi ^a}$. 

It should, however, be noted here that by making use of Gauss's law
\begin{equation}
{\nabla ^2}{\phi ^a} = g\left( {{\rho ^a} - {f^{abc}}{{\bf A}^b} \cdot {{\bf E}^c}} \right), \label{NCYM15}
\end{equation}
we get
\begin{eqnarray}
{\nabla ^2}{\phi ^a} &=& \left( {g{\delta ^{ap}} + {g^2}{f^{abp}}{{\bf A}^b} \cdot {\bf\nabla} \frac{1}{{{\nabla ^2}}} + {g^3}{f^{abc}}{{\bf A}^b} \cdot \nabla \frac{1}{{{\nabla ^2}}}{f^{chp}}{{\bf A}^h} \cdot \nabla \frac{1}{{{\nabla ^2}}}} \right) \nonumber\\
&\times&\left( {{\rho ^p} - {f^{pde}}{{\bf A}^d} \cdot {\bf E}_T^e} \right). \label{NCYM20}
\end{eqnarray}

Next, the corresponding formulation of this theory in the presence of a minimal length is by means of a smeared source \cite{Gaete:2011ka,Gaete:2012yu}. As a consequence, we will take the sources as$\rho ^a  \equiv \rho _1^a  + \rho _2^a  = \rho _{\bar q}^a  + \rho _q^a $, where $\rho _{\bar q}^a({\bf x})  = t_{\bar q}^a e^{{\raise0.5ex\hbox{$\scriptstyle \theta $}\kern-0.1em/\kern-0.15em
\lower0.25ex\hbox{$\scriptstyle 2$}}\nabla ^2 } \delta ^{\left( 3 \right)} \left( {{\bf x} - {\bf y} \prime } \right)$ and $
\rho _q^a({\bf x}) = t_q^a e^{{\raise0.5ex\hbox{$\scriptstyle \theta $}
\kern-0.1em/\kern-0.15em
\lower0.25ex\hbox{$\scriptstyle 2$}}\nabla ^2 } \delta ^{\left( 3 \right)} \left( {{\bf x} - {\bf y}} \right)$. As in \cite{Bagan:2000nc}, $t_{\bar q}^a$ and $t_q^a$ are the color charges of a heavy antiquark ${\bar q}_i$ and a quark $q_i$ in a normalized color singlet state $  
\left| \Psi  \right\rangle  = N^{ - {\raise0.5ex\hbox{$\scriptstyle 1$}
\kern-0.1em/\kern-0.15em
\lower0.25ex\hbox{$\scriptstyle 2$}}} \left| {q_i } \right\rangle \left| {\bar q_i } \right\rangle$. Hence $
t_q^a t_{\bar q}^a  = {\textstyle{1 \over N}}tr\left( {T^a T^a } \right) =  - C_F$, where the anti-Hermitian generators $T^a$ are in the fundamental representation of SU(N).

Having characterized the sources, we can now compute the expectation value of the energy operator H. Thus, to order $g^2$  and $g^4$, we then easily verify that
\begin{equation}
V=V_1 +V_2, \label{NCYM25}
\end{equation}
where
\begin{equation}
V_1 =  - {g^2}\int {{d^3}x} \left\langle 0 \right|{\rho _1^a}\frac{1}{{{\nabla ^2}}}{\rho _2^a}\left| 0 \right\rangle, \label{NCYM30}
\end{equation}
and
\begin{equation}
V_2=- 3{g^4}f^{abc}f^{chq}\int {{d^3}x} \left\langle 0 \right|{\rho _1^a}\frac{1}{{{\nabla ^2}}}{{\bf A}^b} \cdot \nabla \frac{1}{{{\nabla ^2}}}{{\bf A}^h} \cdot \nabla \frac{1}{{{\nabla ^2}}}{\rho _2^q}\left| 0 \right\rangle. \label{NCYM35}
\end{equation}

The $V_1$ term leads immediately to the result
\begin{equation}
V_1  =  - g^2 C_F \int {\frac{{d^3 k}}{{\left( {2\pi } \right)^3 }}} \frac{{e^{ - \theta {\bf k}^2 } }}{{{\bf k}^2 }}e^{ - i{\bf k} \cdot {\bf r}}  =  - g^2 C_F \frac{1}{{4\pi ^{{\raise0.5ex\hbox{$\scriptstyle 3$}
\kern-0.1em/\kern-0.15em
\lower0.25ex\hbox{$\scriptstyle 2$}}} }}\frac{1}{r}\gamma \left( {{\textstyle{1 \over 2}},{\textstyle{{r^2 } \over {4\theta }}}} \right)
, \label{NCYM45}
\end{equation}
where $|{\bf r}|\equiv |{\bf y}-{\bf y}^\prime|= r$ and $\gamma \left( {{\textstyle{1 \over 2}},{\textstyle{{r^2 } \over {4\theta }}}} \right)$ is the lower incomplete Gamma function defined by the following integral representation
\begin{equation}
\gamma \left(\, \frac{a}{b}\, \right) \equiv \int_0^x {\frac{{du}}{u}}\,
u^{{a \mathord{\left/
{\vphantom {a b}} \right.
\kern-\nulldelimiterspace} b}}\, e^{ - u}.
\label{NCYM45b}
\end{equation}
Hence we see that the term of order $g^2$ is just the regular Coulomb energy at the origin due to the color charges of the quarks.

We now come to the $V_2$ term, which is given by
\begin{equation}
V_2  = 3g^4 C_A C_F \int {\frac{{d^3 k}}{{\left( {2\pi } \right)^3 }}} \frac{{e^{ - \theta {\bf k}^2 } }}{{{\bf k}^2 }}e^{ - i{\bf k} \cdot {\bf r}} {\cal I}({\bf k}), \label{NCYM50}
\end{equation}
where
\begin{equation}
{\cal I}({\bf k})=\int {\frac{{d^3 p}}{{\left( {2\pi } \right)^3 }}} \frac{1}{{2|{\bf p}|\left( {{\bf p} - {\bf k}} \right)^2 }}\left( {1 - \frac{{\left( {{\bf p} \cdot {\bf k}} \right)^2 }}{{{\bf p}^2 {\bf k}^2 }}} \right). \label{NCYM55}
\end{equation}
To get Eq.(\ref{NCYM50}) we have expressed the $A^{ai}$-fields in terms of a normal mode expansion: $    
A^{ai} ({\bf x},t) = \int {\frac{{d^3 p}}{{\sqrt {\left( {2\pi } \right)^3 2w_{\bf p} } }}} \sum\limits_\lambda  {\varepsilon ^i } \left( {{\bf p},\lambda } \right)\left[ {a^a \left( {{\bf p},\lambda } \right)e^{ - ipx}  + a^{\dag a} \left( {{\bf p},\lambda } \right)e^{ipx} } \right]$. Also we have used $    
\left[ {a^a \left( {{\bf p},\lambda } \right),a^{\dag b} \left( {{\bf l},\sigma } \right)} \right] = \delta ^{ab} \delta ^{\lambda \sigma } \delta ^{\left( 3 \right)} \left( {{\bf p} - {\bf l}} \right)$ and $  
\sum\limits_\lambda  {\varepsilon ^i \left( {{\bf k},\lambda } \right)} \varepsilon ^j \left( {{\bf k},\lambda } \right) = \delta ^{ij}  - \frac{{k^i k^j }}{{{\bf k}^2 }}$.  It is worth recalling here that the correction term of order $g^4$ represents an anti-screening effect and makes the interquark potential weaker at short distances, which is in the origin of asymptotic freedom in QCD. Incidentally, it is of interest to note that the anti-screening effect is due to the instantaneous Coulomb interaction of the quarks.

Our next task is to compute the integral (\ref{NCYM55}). In fact, when this integral is performed it is found that by introducing a cutoff $(\Lambda)$ is just sufficient to extract the logarithmic divergence. Thus, the corresponding integration gives ${\cal I}({\bf k}) = \frac{1}{{16\pi ^2 }}\frac{4}{3}\ln \left( {\frac{{{\bf k}^2 }}{{\Lambda ^2 }}} \right)$. This then implies that the integral of Eq. (\ref{NCYM50}) can be cast under the form: 
\begin{equation}
V_2  = g^4 C_F C_A \int {\frac{{d^3 k}}{{\left( {2\pi } \right)^3 }}} \frac{{e^{ - \theta {\bf k}^2 } }}{{{\bf k}^2 }}e^{ - i{\bf k} \cdot {\bf r}} \frac{{12}}{{48\pi ^2 }}\ln \left( {\frac{{{\bf k}^2 }}{{\Lambda ^2 }}} \right). \label{NCYM60}
\end{equation}

Now, we move on to compute the integral (\ref{NCYM60}). To do this it is advantageous to introduce a new auxiliary function (${\cal F}$) \cite{Peter:1997me}:
\begin{equation}
{\cal F}(r,\Lambda ,s) \equiv \Lambda ^{2s} \int {\frac{{d^3 k}}{{\left( {2\pi } \right)^3 }}} \frac{{e^{ - \theta {\bf k}^2 } }}{{\left( {{\bf k}^2 } \right)^{1 + s} }}e^{-i{\bf k} \cdot {\bf r}}, \label{NCYM65}
\end{equation}
such that the $V_2$ term takes the form 

\begin{equation}
V_2  =  - \frac{{12}}{{48\pi ^2 }}g^4 C_F C_A \frac{\partial }{{\partial s}}\left[ {{\cal F}(r,\Lambda ,s)} \right]_{s = 0}. \label{NCYM70}
\end{equation}

We also note that  
\begin{equation}
{\cal F}(r,\Lambda ,s) = \frac{{\Lambda ^{2s} 2^{2s - 2} }}{{\pi ^2 }}\frac{{\Gamma \left( {{\raise0.5ex\hbox{$\scriptstyle 1$}
\kern-0.1em/\kern-0.15em
\lower0.25ex\hbox{$\scriptstyle 2$}} + s} \right)}}{{\Gamma \left( {1 + 2s} \right)}}\frac{1}{r}\int_0^{{\raise0.5ex\hbox{$\scriptstyle {r^2 }$}
\kern-0.1em/\kern-0.15em
\lower0.25ex\hbox{$\scriptstyle {4\theta }$}}} {dy} \frac{1}{{y^{{\raise0.5ex\hbox{$\scriptstyle 1$}
\kern-0.1em/\kern-0.15em
\lower0.25ex\hbox{$\scriptstyle 2$}}} }}\left( {\frac{{r^2 }}{{4y}} - \theta } \right)^s e^{ - y}. \label{NCYM75}
\end{equation}
Hence we see that, at leading order in $\theta$, the function ${\cal F}$ is then 
\begin{eqnarray}
{\cal F}(r,\Lambda ,s) &=&\frac{{\left( {\Lambda r} \right)^{2s} }}{{4\pi ^2 }}\frac{{\Gamma \left( {{\raise0.5ex\hbox{$\scriptstyle 1$}
\kern-0.1em/\kern-0.15em
\lower0.25ex\hbox{$\scriptstyle 2$}} + s} \right)}}{{\Gamma \left( {1 + 2s} \right)}}\frac{1}{r}\gamma \left( {{\raise0.5ex\hbox{$\scriptstyle 1$}
\kern-0.1em/\kern-0.15em
\lower0.25ex\hbox{$\scriptstyle 2$}} - s,{\raise0.5ex\hbox{$\scriptstyle {r^2 }$}
\kern-0.1em/\kern-0.15em
\lower0.25ex\hbox{$\scriptstyle {4\theta }$}}} \right) \nonumber\\
&-& \frac{{\theta s\left( {\Lambda r} \right)^{2s} }}{{\pi ^2 }}\frac{{\Gamma \left( {{\raise0.5ex\hbox{$\scriptstyle 1$}
\kern-0.1em/\kern-0.15em
\lower0.25ex\hbox{$\scriptstyle 2$}} + s} \right)}}{{\Gamma \left( {1 + 2s} \right)}}\frac{1}{{r^3 }}\gamma \left( {{\raise0.5ex\hbox{$\scriptstyle 3$}
\kern-0.1em/\kern-0.15em
\lower0.25ex\hbox{$\scriptstyle 2$}} - s,{\raise0.5ex\hbox{$\scriptstyle {r^2 }$}
\kern-0.1em/\kern-0.15em
\lower0.25ex\hbox{$\scriptstyle {4\theta }$}}} \right). \label{NCYM80}
\end{eqnarray}
With this, we can write (\ref{NCYM70}) also as
\begin{eqnarray}
V_2  =  &-& \frac{1}{{2\pi ^{{\raise0.5ex\hbox{$\scriptstyle 3$}
\kern-0.1em/\kern-0.15em
\lower0.25ex\hbox{$\scriptstyle 2$}}} }}\frac{{g^4 }}{{\pi ^2 }}\frac{{12}}{{48}}C_A C_F \frac{1}{r}\ln \left( {\Lambda r} \right)\gamma \left( {{\raise0.5ex\hbox{$\scriptstyle 1$}
\kern-0.1em/\kern-0.15em
\lower0.25ex\hbox{$\scriptstyle 2$}},{\raise0.5ex\hbox{$\scriptstyle {r^2 }$}
\kern-0.1em/\kern-0.15em
\lower0.25ex\hbox{$\scriptstyle {4\theta }$}}} \right) \nonumber\\
&+& \frac{\theta }{{\pi ^{{\raise0.5ex\hbox{$\scriptstyle 3$}
\kern-0.1em/\kern-0.15em
\lower0.25ex\hbox{$\scriptstyle 2$}}} }}\frac{{g^4 }}{{\pi ^2 }}\frac{{12}}{{48}}C_A C_F \frac{1}{{r^3 }}\gamma \left( {{\raise0.5ex\hbox{$\scriptstyle 3$}
\kern-0.1em/\kern-0.15em
\lower0.25ex\hbox{$\scriptstyle 2$}},{\raise0.5ex\hbox{$\scriptstyle {r^2 }$}
\kern-0.1em/\kern-0.15em
\lower0.25ex\hbox{$\scriptstyle {4\theta }$}}} \right). \label{NCYM85}
\end{eqnarray}
We draw attention to the fact that the previous result is finite at order $\theta$. In fact, one comment is pertinent in this context. Since we have considered the quantization of the full theory, without expanding in powers of $\theta$ and no truncation was made, we get a finite result. The $\theta$-expansion of the function $\cal F$ is only for reasons of calculation. This has nothing to do with the $\theta$-expansion through the Moyal $\star$-product, as we have explained in \cite{Gaete:2011ka}.

We now want to consider the $(g^4)$ screening contribution to the potential, which is due to the exchange of transverse gluons. This effect makes the interquark potential stronger at short distances. With this in mind, from perturbation theory, we start by writing
\begin{eqnarray}
V_2^ *   &=& 2g^4 f_{abc} f_{def} \sum\limits_{n = 2 \ gluon} {\frac{1}{{E_n }}} \int {d^3 x} \int {d^3 w} \left\langle 0 \right|\rho _2^a \frac{1}{{\nabla ^2 }}{\bf A}^b \cdot {\bf E}_T^c \left| n \right\rangle _{\bf x} \nonumber\\
&\times&\left\langle n \right|\rho _1^d \frac{1}{{\nabla ^2 }}{\bf A}^e \cdot {\bf E}_T^f \left| 0 \right\rangle _{\bf w},\label{NCYM90} 
\end{eqnarray}
where, by construction, $\left\langle 0 \right|\rho _2^a \frac{1}{{\nabla ^2 }}{\bf A}^b \cdot {\bf E}_T^c \left| n \right\rangle$ is the matrix element in the basis of states in which the non perturbated Hamiltonian is diagonal. In order to evaluate Eq. (\ref{NCYM90}), we note that the intermediate state $\left| n \right\rangle$ must contains a pair of transverse gluons, since the terms $ {\bf A}^b \cdot {\bf E}_T^c $ must create and destroy dynamical gluon pairs. Taking this remark into account, we may write two gluon states as 
\begin{eqnarray}
\sum\limits_{n = 2 \ gluon} {\left| n \right\rangle } \left\langle n \right| &=& \frac{1}{2}\sum\limits_{kl}\sum\limits_{\lambda \sigma } {\int {d^3 k} } \int {d^3 } l\ a^{\dag e} \left( {{\bf  k},\lambda } \right)a^{\dag f} \left( {{\bf    l},\sigma } \right)\left| 0 \right\rangle \nonumber\\
&\times& \left\langle 0 \right|a^f \left( {{\bf  l},\sigma } \right)a^e \left( {{\bf  k},\lambda } \right). 
\label{NCYM90b}
\end{eqnarray}
Substitution of Eq. (\ref{NCYM90b}) into Eq. (\ref{NCYM90}) and following our earlier procedure, the $ V_2^ *$ term reduces to
\begin{equation}
V_2^ *   =  - C_A C_F g^4 \int {\frac{{d^3 k}}{{\left( {2\pi } \right)^3 }}} \frac{{e^{ - \theta {\bf k}^2 } }}{{{\bf k}^4 }}e^{i{\bf k} \cdot {\bf r}} {\cal I}\left( {\bf k} \right), \label{NCYM95}
\end{equation}
where
\begin{equation}
{\cal I}\left( {\bf k} \right) = \int {\frac{{d^3 l}}{{\left( {2\pi } \right)^3 }}} \frac{{\left( {|{\bf l}| - |{\bf l} - {\bf k}|} \right)^2 }}{{4|{\bf l}{\bf |}|{\bf l} - {\bf k}|\left( {|{\bf l}| + |{\bf l} - {\bf k}|} \right)}}\left\{ {1 + \frac{{\left[ {\left( {{\bf l} - {\bf k}} \right) \cdot {\bf l}} \right]^2 }}{{{\bf l}^2 \left( {{\bf l} - {\bf k}} \right)^2 }}} \right\}. \label{NCYM100}
\end{equation}

Integrating now over {\bf k}, one then obtains ${\cal I}\left( {\bf k} \right) = \frac{1}{{48\pi ^2 }}\ln \left( {\frac{{{\bf k}^2 }}{{\Lambda ^2 }}} \right)$. As a consequence, the $V_2^ *$ term becomes
\begin{equation}
V_2^ *   =  - g^4 C_F C_A \int {\frac{{d^3 k}}{{\left( {2\pi } \right)^3 }}\frac{{e^{ - \theta {\bf k}^2 } }}{{{\bf k}^2 }}} e^{ - i{\bf k} \cdot {\bf r}} \frac{1}{{48\pi ^2 }}\ln \left( {\frac{{{\bf k}^2 }}{{\Lambda ^2 }}} \right). \label{NCYM105}
\end{equation}
It is straightforward to see that this integral is exactly the one obtained in expression (\ref{NCYM60}).

By putting together Eqs. (\ref{NCYM45}), (\ref{NCYM60}) and (\ref{NCYM105}), we obtain for the total interquak potential

\begin{eqnarray}
V = - g^2 C_F \int {\frac{{d^3 k}}{{\left( {2\pi } \right)^3 }}} \frac{{e^{ - \theta k^2 } }}{{k^2 }}\left\{ {1 + g^2 C_A \frac{{11}}{{48\pi ^2 }}\ln \left( {\frac{{\Lambda ^2 }}{{k^2 }}} \right)} \right\}e^{ - ik \cdot r}. \label{NCYM110}
\end{eqnarray}
By means of the function ${\cal F}$ we evaluate the interquark potential in position space. Accordingly, one finds
\begin{eqnarray}
V = &-& \frac{{g^2 C_F }}{{4\pi ^{{\raise0.5ex\hbox{$\scriptstyle 3$}
\kern-0.1em/\kern-0.15em
\lower0.25ex\hbox{$\scriptstyle 2$}}} }}\frac{1}{r}\left( {1 + g^2 C_A \frac{{11}}{{24\pi ^2 }}\ln \left( {\Lambda r} \right)} \right)\gamma \left( {{\raise0.5ex\hbox{$\scriptstyle 1$}
\kern-0.1em/\kern-0.15em
\lower0.25ex\hbox{$\scriptstyle 2$}},{\raise0.5ex\hbox{$\scriptstyle {r^2 }$}
\kern-0.1em/\kern-0.15em
\lower0.25ex\hbox{$\scriptstyle {4\theta }$}}} \right) \nonumber\\
&+& \theta g^4 C_F C_A \frac{{11}}{{48\pi ^{{\raise0.5ex\hbox{$\scriptstyle 7$}
\kern-0.1em/\kern-0.15em
\lower0.25ex\hbox{$\scriptstyle 2$}}} }}\frac{1}{{r^3 }}\gamma \left( {{\raise0.5ex\hbox{$\scriptstyle 3$}
\kern-0.1em/\kern-0.15em
\lower0.25ex\hbox{$\scriptstyle 2$}},{\raise0.5ex\hbox{$\scriptstyle {r^2 }$}
\kern-0.1em/\kern-0.15em
\lower0.25ex\hbox{$\scriptstyle {4\theta }$}}} \right). \label{NCYM115}
\end{eqnarray}
An immediate consequence of this is that for $\theta=0$ one obtains the known interquark potential at lowest order in $g$ \cite{Drell:1981gu}.

Interestingly, it is observed that the term proportional to $\theta$ is ultraviolet finite. Another important point to be mentioned in our discussion comes from the explicit appearance of the cutoff $\Lambda$ in (\ref{NCYM115}). Evidently, at distances higher than 
${\raise0.7ex\hbox{$1$} \!\mathord{\left/
 {\vphantom {1 \Lambda }}\right.\kern-\nulldelimiterspace}
\!\lower0.7ex\hbox{$\Lambda $}}$ the interaction potential (\ref{NCYM115}) is convergent. Nevertheless, at distances lower than
${\raise0.7ex\hbox{$1$} \!\mathord{\left/
 {\vphantom {1 \Lambda }}\right.\kern-\nulldelimiterspace}
\!\lower0.7ex\hbox{$\Lambda $}}$ the static potential is divergent. This result immediately shows that the interaction energy concept is not suitable in this region. In other words, 
${\raise0.7ex\hbox{$1$} \!\mathord{\left/
 {\vphantom {1 \Lambda }}\right.\kern-\nulldelimiterspace}
\!\lower0.7ex\hbox{$\Lambda $}}$ naturally defines a critical distance below which the field regime prevails, that is, the distance scale for which we can no longer describe the interaction by using the potential or force concept, but only by the field one. Also, (below $ 
{\raise0.7ex\hbox{$1$} \!\mathord{\left/
 {\vphantom {1 \Lambda }}\right.\kern-\nulldelimiterspace}
\!\lower0.7ex\hbox{$\Lambda $}}$) our description must be supplemented by degrees of freedom that are truncated below the cutoff $\Lambda$.

\section{Gluodynamics in curved space-time in the presence of a minimal length}

We now pass to the calculation of the interquark potential for gluodynamics in curved space-time \cite{Gaete:2007zn}. In other words, we wish to explore the effects of including a minimal length on the nature of the potential. The corresponding theory is governed by the Lagrangian density:
\begin{equation}
{\cal L} = \frac{{\left| {\varepsilon _V } \right|}}{{m^2
}}\frac{1}{2}e^{{\raise0.7ex\hbox{$\chi $} \!\mathord{\left/
 {\vphantom {\chi  2}}\right.\kern-\nulldelimiterspace}
\!\lower0.7ex\hbox{$2$}}} \left( {\partial _\mu  \chi } \right)^2  +
\left| {\varepsilon _V } \right|e^\chi  \left( {1 - \chi } \right) 
- e^\chi  \left( {1 - \chi } \right)\frac{1}{4}F_{\mu \nu }^a F^{a\mu
\nu }, \label{Gluon5}
\end{equation}
where the real scalar field $\chi$ of mass $m$ represents the
dilaton, and $- \left| {\varepsilon _V } \right|$ is the vacuum
energy density. Thus, after to retain only the leading quadratic term and integrating over the $\chi$-field, we arrive at the following effective Lagrangian 
\begin{equation}
{\cal L}_{eff}  =  - \frac{1}{4}F_{\mu \nu }^a \left( {1 + \frac{{m^2 }}{\Delta }} \right)F^{a\mu \nu} + |\varepsilon _V |, \label{Gluon10}
\end{equation}
we have skipped all the technical details and refer to \cite{Gaete:2007zn} for them.

Now, proceeding as before, the Hamiltonian of the theory is given by
\begin{equation}
H =  - \frac{1}{2}\int {d^3 x}\ \phi ^a \left( {\nabla ^2  - m^2 } \right)\phi ^a . \label{Gluon15}
\end{equation}
In analogy to (\ref{NCYM20}) we have 
\begin{eqnarray}
\left( {\nabla ^2  - m^2 } \right)\phi ^a  &=& \left( {g\delta ^{ap}  + g^2 f^{abp} {\bf A}^b  \cdot \nabla \frac{1}{{\nabla ^2 }} + g^3 f^{abc} {\bf A}^b  \cdot \nabla \frac{1}{{\nabla ^2 }}f^{chp} {\bf A}^h  \cdot \nabla \frac{1}{{\nabla ^2 }}} \right) \nonumber\\
&\times&\left( {\rho ^p  - f^{pde} {\bf A}^d  \cdot \left[ {\frac{{\nabla ^2  - m^2 }}{{\nabla ^2 }}} \right]{\bf E}_T^e } \right). \label{Gluon20}
\end{eqnarray}

We are now in a position to evaluate the interquark potential for gluodynamics in curved space-time. In this case, we see that the expectation value $\left\langle 0 \right|H\left| 0 \right\rangle$ reads:
\begin{equation}
V = V_1  + V_2. \label{Gluon25}
\end{equation}
The term $V_1$ ($V_1\equiv V_1^{\left( 1 \right)} + V_1^{\left( 2 \right)}$) is now given by
\begin{equation}
V_1^{\left( 1 \right)}  = g^2 \int {d^3 x} \left\langle 0 \right|\rho _1^a \frac{1}{{\nabla ^2  - m^2 }}\rho _2^a \left| 0 \right\rangle, \label{Gluon25b}
\end{equation}
and
\begin{equation}
V_1^{\left( 2 \right)}  = 3g^4 f^{abc} f^{chq} \int {d^3 x} \left\langle 0 \right|\rho _1^a \frac{1}{{\nabla ^2  - m^2 }}{\bf A}^b  \cdot \nabla \frac{1}{{\nabla ^2 }}{\bf A}^h  \cdot \nabla \frac{1}{{\nabla ^2 }}\rho _2^q \left| 0 \right\rangle. \label{Gluon25c}
\end{equation}

As was showed in \cite{Gaete:2011ka}, the $V_1^{\left( 1 \right)}$ term reduces to
\begin{equation}
V_1^{\left( 1 \right)}  =  - g^2 C_F \frac{{e^{m^2 \theta } }}{{4\pi }}\frac{1}{r}\left[ {e^{ - mr}  - \frac{1}{{\sqrt \pi  }}\int_{{\raise0.5ex\hbox{$\scriptstyle {r^2 }$}
\kern-0.1em/\kern-0.15em
\lower0.25ex\hbox{$\scriptstyle {4\theta }$}}}^\infty  {d\xi \xi ^{ - {\raise0.5ex\hbox{$\scriptstyle 1$}
\kern-0.1em/\kern-0.15em
\lower0.25ex\hbox{$\scriptstyle 2$}}} } e^{ - \xi  - {\raise0.5ex\hbox{$\scriptstyle {mr^2 }$}
\kern-0.1em/\kern-0.15em
\lower0.25ex\hbox{$\scriptstyle {4\xi }$}}} } \right].  \label{Gluon30}
\end{equation}
While, following the same steps as those of the preceding section, the $V_1^{\left( 2 \right)}$ term turns out to be
\begin{equation}
V_1^{\left( 2 \right)}  =  - g^4 \frac{{12}}{{48\pi ^2 }}C_F C_A \frac{\partial }{{\partial s}}\left[ {\Lambda ^{2s} \int {\frac{{d^3 k}}{{\left( {2\pi } \right)^3 }}\frac{{e^{ - \theta k^2 } }}{{\left( {{\bf k}^2  + m^2 } \right)}}\frac{1}{{\left( {{\bf k}^2 } \right)^s }}e^{ - i{\bf k} \cdot {\bf r}} } } \right]_{s = 0}.  \label{Gluon35}
\end{equation} 
For our purposes it is sufficient to retain the leading quadratic term in $\bf k$. Thus, the $V_1^{\left( 2 \right)}$ term simplifies to
\begin{equation}
V_1^{\left( 2 \right)}  =  - g^4 \frac{{12}}{{48\pi ^2 }}C_F C_A \frac{\partial }{{\partial s}}{\cal F}\left( {r,\Lambda ,s} \right), \label{Gluon40} 
\end{equation}
where ${\cal F}$ was defined in (\ref{NCYM65}).

Now we turn our attention to the $V_2$ term, which this time is expressed as, 
\begin{eqnarray}
V_2  &=& 2g^4 \sum\limits_{n = 2 \ gluon} {\frac{1}{{E_n }}} f^{abc} f^{def} \int {d^3 x} \int {d^3 w} \nonumber\\
&\times&\left\langle 0 \right|\rho _2^a \frac{1}{{\nabla ^2  - m^2 }}{\bf A}^b  \cdot \left( {\frac{{\nabla ^2  - m^2 }}{{\nabla ^2 }}} \right){\bf E}_T^c \left| n \right\rangle \nonumber\\
&\times&\left\langle n \right|\rho _1^d \frac{1}{{\nabla ^2  - m^2 }}{\bf A}^e  \cdot \left( {\frac{{\nabla ^2  - m^2 }}{{\nabla ^2 }}} \right){\bf E}_T^f \left| 0 \right\rangle. \label{Gluon45}
\end{eqnarray}
Unlike the previous subsection, this latter term consists of two parts, namely:
\begin{eqnarray}
V_2^{\left( {1} \right)}  &=& g^4 f^{abc} f^{def} \int {d^3 x} \int {d^3 w} \int {d^3 k} \int {d^3 l} \sum\limits_{m,n}\sum\limits_{\lambda ,\sigma } {\frac{1}{{E_n }}} \nonumber\\
&\times&\left\langle 0 \right|\rho _2^a \frac{1}{{\nabla ^2  - m^2 }}{\bf A}^b  \cdot {\bf E}_T^c a^{\dag m} \left( {{\bf k},\lambda } \right)a^{\dag n} \left( {{\bf l},\sigma } \right)\left| 0 \right\rangle _{\bf x} \nonumber\\
&\times&\left\langle 0 \right|a^n \left( {{\bf l},\sigma } \right)a^m \left( {{\bf k},\lambda } \right)\rho _1^d \frac{1}{{\nabla ^2  - m^2 }}{\bf A}^e  \cdot {\bf E}_T^f \left| 0 \right\rangle _{\bf w}, \label{Gluon50}
\end{eqnarray}
and
\begin{eqnarray}
V_2^{\left( 2 \right)}  &=& m^4 g^4 f^{abc} f^{def} \int {d^3 x} \int {d^3 w} \int {d^3 k} \int {d^3 l} \sum\limits_{m,n}\sum\limits_{\lambda ,\sigma } {\frac{1}{{E_n }}} 
\nonumber\\
&\times&\left\langle 0 \right|\rho _2^a \frac{1}{{\nabla ^2  - m^2 }}{\bf A}^b  \cdot \frac{1}{{\nabla ^2 }}{\bf E}_T^c a^{\dag m} \left( {{\bf k},\lambda } \right)a^{\dag n} \left( {{\bf l},\sigma } \right)\left| 0 \right\rangle _{\bf x} \nonumber\\
&\times&\left\langle 0 \right|a^n \left( {{\bf l},\sigma } \right)a^m \left( {{\bf k},\lambda } \right)\rho _1^d \frac{1}{{\nabla ^2  - m^2 }}{\bf A}^e  \cdot \frac{1}{{\nabla ^2 }}{\bf E}_T^f \left| 0 \right\rangle_{\bf w} .\label{Gluon55}
\end{eqnarray}
Following our earlier procedure, these expressions can be conveniently rewritten as 
\begin{equation}
V_2^{\left( 1 \right)}  =  - g^4 C_F C_A \int {\frac{{d^3 k}}{{\left( {2\pi } \right)^3 }}} \frac{{e^{ - \theta {\bf k}^2 } }}{{\left( {{\bf k}^2  + m^2 } \right)^2 }}\frac{1}{{48\pi ^2 }}\ln \left( {\frac{{{\bf k}^2 }}{{\Lambda ^2 }}} \right)e^{i{\bf k} \cdot {\bf r}}, \label{Gluon60}
\end{equation}
and
\begin{equation}
V_2^{\left( 2 \right)}  =  - m^4 g^4 C_F C_A \int {\frac{{d^3 k}}{{\left( {2\pi } \right)^3 }}} \frac{{e^{ - \theta {\bf k}^2 } }}{{\left( {{\bf k}^2  + m^2 } \right)^2 {\bf k}^2 }}e^{i{\bf k} \cdot {\bf r}} . \label{Gluon65}
\end{equation}
Again, since we are interested in estimating the lowest-order correction to the interquark potential, we will retain only the leading quadratic terms in the expressions (\ref{Gluon60}) and (\ref{Gluon65}), namely,
\begin{equation}
V_2^{\left( 1 \right)}  =  - g^4 C_F C_A \int {\frac{{d^3 k}}{{\left( {2\pi } \right)^3 }}} \frac{{e^{ - \theta {\bf k}^2 } }}{{{\bf k}^4 }}\frac{1}{{48\pi ^2 }}\ln \left( {\frac{{{\bf k}^2 }}{{\Lambda ^2 }}} \right)e^{i{\bf k} \cdot {\bf r}}, \label{Gluon70} 
\end{equation}
and
\begin{equation}
V_2^{\left( 2 \right)}  =  - m^2 g^4 C_F C_A \int {\frac{{d^3 k}}{{\left( {2\pi } \right)^3 }}} \frac{{e^{ - \theta {\bf k}^2 } }}{{{\bf k}^4 }}e^{i{\bf k} \cdot {\bf r}}. \label{Gluon75}
\end{equation}
In position space we then have
\begin{equation}
V_2^{\left( 1 \right)}  = g^4 \frac{1}{{48\pi ^2 }}C_F C_A \frac{\partial }{{\partial s}}\left[ {{\cal F}\left( {r,\Lambda ,s} \right)} \right]_{s = 0}, \label{Gluon80}
\end{equation}
and
\begin{equation}
V_2^{\left( 2 \right)}  = \frac{{m^2 g^4 }}{{32\sqrt 2 \pi ^{{\raise0.5ex\hbox{$\scriptstyle 7$}
\kern-0.1em/\kern-0.15em
\lower0.25ex\hbox{$\scriptstyle 2$}}} }}\left\{ {r\  \gamma \left( {{\raise0.5ex\hbox{$\scriptstyle 1$}
\kern-0.1em/\kern-0.15em
\lower0.25ex\hbox{$\scriptstyle 2$}},{\raise0.5ex\hbox{$\scriptstyle {r^2 }$}
\kern-0.1em/\kern-0.15em
\lower0.25ex\hbox{$\scriptstyle {4\theta }$}}} \right) + 2\sqrt \theta  e^{ - {\raise0.5ex\hbox{$\scriptstyle {r^2 }$}
\kern-0.1em/\kern-0.15em
\lower0.25ex\hbox{$\scriptstyle {4\theta }$}}}  + \frac{{2\theta }}{r}\gamma \left( {{\raise0.5ex\hbox{$\scriptstyle 1$}
\kern-0.1em/\kern-0.15em
\lower0.25ex\hbox{$\scriptstyle 2$}},{\raise0.5ex\hbox{$\scriptstyle {r^2 }$}
\kern-0.1em/\kern-0.15em
\lower0.25ex\hbox{$\scriptstyle {4\theta }$}}} \right)} \right\}. \label{Gluon85}
\end{equation}

Finally, by putting together equations (\ref{Gluon30}), (\ref{Gluon40}), (\ref{Gluon80}), and (\ref{Gluon85}), we obtain for the total interquark potential
\begin{eqnarray}
V &=&  - g^2 C_F \frac{{e^{m^2 \theta } }}{{4\pi }}\frac{1}{r}\left\{ {e^{ - mr}  - \frac{1}{{\sqrt \pi  }}\int_{{\raise0.5ex\hbox{$\scriptstyle {r^2 }$}
\kern-0.1em/\kern-0.15em
\lower0.25ex\hbox{$\scriptstyle {4\theta }$}}}^\infty  {d\xi } \xi ^{ - {\raise0.5ex\hbox{$\scriptstyle 1$}
\kern-0.1em/\kern-0.15em
\lower0.25ex\hbox{$\scriptstyle 2$}}} e^{ - \xi  - m{\raise0.5ex\hbox{$\scriptstyle {r^2 }$}
\kern-0.1em/\kern-0.15em
\lower0.25ex\hbox{$\scriptstyle {4\xi }$}}} } \right\}  \nonumber\\
&-& \frac{{g^4 C_A C_F }}{{16\pi ^{{\raise0.5ex\hbox{$\scriptstyle 7$}
\kern-0.1em/\kern-0.15em
\lower0.25ex\hbox{$\scriptstyle 2$}}} }}\frac{1}{r}\left\{ {\frac{{11}}{6}\ln (\Lambda r) - \frac{{m^2 \theta }}{{\sqrt 2 }}} \right\}\gamma \left( {{\raise0.5ex\hbox{$\scriptstyle 1$}
\kern-0.1em/\kern-0.15em
\lower0.25ex\hbox{$\scriptstyle 2$}},{\raise0.5ex\hbox{$\scriptstyle {r^2 }$}
\kern-0.1em/\kern-0.15em
\lower0.25ex\hbox{$\scriptstyle {4\theta }$}}} \right) + \frac{{11g^4 C_A C_F }}{{48\pi ^{{\raise0.5ex\hbox{$\scriptstyle 7$}
\kern-0.1em/\kern-0.15em
\lower0.25ex\hbox{$\scriptstyle 2$}}} }}\frac{\theta }{{r^3 }}\gamma \left( {{\raise0.5ex\hbox{$\scriptstyle 1$}
\kern-0.1em/\kern-0.15em
\lower0.25ex\hbox{$\scriptstyle 2$}},{\raise0.5ex\hbox{$\scriptstyle {r^2 }$}
\kern-0.1em/\kern-0.15em
\lower0.25ex\hbox{$\scriptstyle {4\theta }$}}} \right) \nonumber\\
&+& \frac{{g^4 m^2 C_A C_F }}{{16\sqrt 2 \pi ^{{\raise0.5ex\hbox{$\scriptstyle 7$}
\kern-0.1em/\kern-0.15em
\lower0.25ex\hbox{$\scriptstyle 2$}}} }}\sqrt \theta  e^{ - {\raise0.5ex\hbox{$\scriptstyle {r^2 }$}
\kern-0.1em/\kern-0.15em
\lower0.25ex\hbox{$\scriptstyle {4\theta }$}}}  + \frac{{g^4 m^2 C_A C_F }}{{32\sqrt 2 \pi ^{{\raise0.5ex\hbox{$\scriptstyle 7$}
\kern-0.1em/\kern-0.15em
\lower0.25ex\hbox{$\scriptstyle 2$}}} }}r\  \gamma \left( {{\raise0.5ex\hbox{$\scriptstyle 1$}
\kern-0.1em/\kern-0.15em
\lower0.25ex\hbox{$\scriptstyle 2$}},{\raise0.5ex\hbox{$\scriptstyle {r^2 }$}
\kern-0.1em/\kern-0.15em
\lower0.25ex\hbox{$\scriptstyle {4\theta }$}}} \right). \label{Gluon90}
\end{eqnarray}
Here, in contrast to our previous analysis \cite{Gaete:2007zn}, the coefficient of the linear term ("string tension") is ultraviolet finite. This result display a marked qualitative departure from its commutative counterpart \cite{Gaete:2007zn}. However, one easily verify that in the limit $m \to 0$ the confinement term vanishes, as it should be. Furthermore, it is straightforward to see that the terms proportional to $g^2$ and to $\theta$ are all ultraviolet finite. In addition, the cutoff $\Lambda$ appears again in the above expression. In this regard we refer to our comments in the previous subsection.

\section{Final remarks}

To conclude, within stationary perturbation theory, we have studied the confinement versus screening issue for gluodynamics in curved space-time in the presence of a minimal length. We have obtained that the static potential is the sum of a Yukawa-type and a linear potential, leading to the confinement of static charges. More interestingly, it was shown that the coefficient of the linear term is ultraviolet finite, where the new quantum of length was crucial to obtain this result. At this point, we would like to remark that our model for gluodynamics is an effective description that comes out upon integration over the dilaton field, whose excitation is massive. We also draw attention to the role played by dilaton in yielding confinement: its mass contribute linearly to the string tension.

Also, from (\ref{NCYM115}) it is easy see that the introduction of a minimal length preserves the asymptotically free behavior at short distance characterizing the non-Abelian character of the strong interactions.

As a final remark it should be mentioned that a correct identification of physical degrees of freedom is a fundamental ingredient for understanding the physics hidden in gauge theories. This then implies that, once the identification has been made, the computation of the potential is achieved by means of Gauss's law.

\ack
I would like to thank Euro Spallucci and Jos\'e A. Helay\"el-Neto for useful discussions. This work was partially supported by Fondecyt (Chile) Grant $N^{\underline 0}$ 1130426.\\

\end{document}